\documentclass[5p,times,12pt,twocolumn]{elsarticle}
\usepackage{tabularx}
\usepackage[usenames, dvipsnames]{color}

\usepackage{amssymb}

\usepackage{lineno,hyperref}

\usepackage{graphicx}
\usepackage{dcolumn}
\usepackage{csquotes}
\usepackage{braket}
\usepackage{bm}
\usepackage{amsmath}

\begin{document}

\begin{frontmatter}

\title{A Search for Possible Long Range Spin Dependent Interactions of the Neutron From Exotic Vector Boson Exchange}

\author[auth1]{C.~Haddock\corref{cor1}}
\address[auth1]{Nagoya University, Furocho, Chikusa Ward, Nagoya, Aichi Prefecture 464-0814, Japan}
\cortext[cor1]{Principal corresponding author}
\ead{cchaddoc@phi.phys.nagoya-u.ac.jp}

\author[auth2]{J.~Amadio}
\address[auth2]{Gettysburg College, 300 N Washington St, Gettysburg, PA 17325, USA}

\author[auth3]{E.~Anderson}
\address[auth3]{Physics Department, Indiana University, Bloomington, Indiana 47408, USA.}

\author[auth4]{L.~Barr\'{o}n-Palos}
\address[auth4]{Instituto de F\`{i}sica, Universidad Nacional Aut\`{o}noma de M\'{e}xico, Apartado Postal 20-364, 01000, M\'{e}xico}

\author[auth2]{B.~Crawford}

\author[auth5]{C.~Crawford}
\address[auth5]{University of Kentucky, Lexington, KY 40506, USA}

\author[auth6]{D.~Esposito}
\address[auth6]{University of Dayton, 300 College Park, Dayton, OH 45469, USA}

\author[auth3]{W.~Fox}

\author[auth7]{I.~Francis}
\address[auth7]{612 S Mitchell St Bloomington, Indiana 47401, USA}

\author[auth8]{J.~Fry}
\address[auth8]{University of Virginia, Charlottesville, VA 22903, USA}

\author[auth9]{H.~Gardiner}
\address[auth9]{Louisiana State University, Baton Rouge, LA 70803, USA}

\author[auth11]{A.~Holley}
\address[auth11]{Tennessee Tech University, 1 William L Jones Dr, Cookeville, TN 38505, USA}

\author[auth3]{K.~Korsak}

\author[auth12]{J.~Lieffers}
\address[auth12]{Embry-Riddle Aeronautical University, 600 S Clyde Morris Blvd, Daytona Beach, FL 32114, USA}

\author[auth2]{S.~Magers}

\author[auth4]{M.~Maldonado-Vel\'{a}zquez}

\author[auth16]{D.~Mayorov}
\address[auth16]{Los Alamos National Lab, Los Alamos, NM 87545, USA}

\author[auth13]{J.~S.~Nico}
\address[auth13]{National Institute of Standards and Technology, 100 Bureau Dr, Gaithersburg, MD 20899, USA}

\author[auth1]{T.~Okudaira}

\author[auth14]{C.~Paudel}
\address[auth14]{Georgia State University, 29 Peachtree Center Avenue, Atlanta, GA 30303, USA}

\author[auth15]{S.~Santra}
\address[auth15]{Bhabha Atomic Research Centre, Trombay, Mumbai, Maharashtra 400085, India}

\author[auth14]{M.~Sarsour}

\author[auth1]{H.~M.~Shimizu}

\author[auth3]{W.~M.~Snow}

\author[auth5]{A.~Sprow}

\author[auth3]{K.~Steffen}

\author[auth10]{H.~E.~Swanson}
\address[auth10]{University of Washington, Seattle, WA 98105, USA}

\author[auth16]{F.~Tovesson}

\author[auth3]{J.~Vanderwerp}

\author[auth2]{P.~A.~Yergeau}





\begin{abstract}
We present a search for possible spin dependent interactions of the neutron with matter through exchange of spin 1 bosons with axial vector couplings as envisioned in possible extensions of the Standard Model. This was sought using a slow neutron polarimeter that passed transversely polarized slow neutrons by unpolarized slabs of material arranged so that interactions would tilt the plane of polarization and develop a component along the neutron momentum. The result for the rotation angle, $\phi^{\prime} = [2.8 \pm 4.6 (stat.) \pm 4.0(sys.)] \times 10^{-5}$\,rad/m is consistent with zero. This result improves the upper bounds on the neutron-matter coupling $g_{A}^{2}$ by about three orders of magnitude for force ranges in the mm\,-\,$\mu$m regime.
\end{abstract}

 \begin{keyword}
 \texttt{NSR}\sep neutron\sep WISP\sep fundamental symmetries
 \end{keyword}

\end{frontmatter}


\section{Introduction}

The possible existence of new interactions in nature with ranges of mesoscopic scale (millimeters to microns), corresponding to exchange boson masses in the 1\,meV to 1\,eV range and with very weak couplings to matter has begun to attract renewed scientific attention. Particles which might act as the mediators are sometimes referred to generically as WISPs (Weakly-Interacting sub-eV Particles)\,\cite{annurev.nucl.012809.104433, ComRenPhys.12.755} in recent theoretical literature. Many theories beyond the Standard Model, including string theories, possess extended symmetries which, when broken at a high energy scale, lead to weakly-coupled light particles with relatively long-range such as axions, arions, familons, and Majorons\,\cite{PhysRevD.81.123530, PDG14}. The well-known Goldstone theorem in quantum field theory guarantees that the spontaneous breaking down of a continuous symmetry at scale $M$ leads to a massless pseudoscalar mode with weak couplings to massive fermions $m$ of order $g=m/M$. The mode can then acquire a light mass (thereby becoming a pseudo-Goldstone boson) of order $m_{boson}=\Lambda^{2}/M$ if there is also an explicit breaking of the symmetry at scale $\Lambda$\,\cite{PhysRevLett.29.1698}. New axial-vector bosons such as paraphotons\,\cite{PhysRevLett.94.151802} and extra Z bosons\,\cite{PhysRevD.68.035012} appear in certain gauge theories beyond the Standard Model. Several theoretical attempts to explain dark matter and dark energy also produce new weakly-coupled long-range interactions. The fact that the dark energy density of order (1\,meV)${^4}$ corresponds to a length scale of $\sim$ 100\,$\mu$m also encourages searches for new phenomena on this scale\,\cite{Ade09}.

A general classification of two-body interactions between nonrelativistic massive spin $1/2$ fermions from the single exchange of a spin $0$ or spin $1$ boson assuming only rotational invariance\,\cite{Dob06} reveals sixteen operator structures involving the spins, momenta, interaction range, and various possible couplings of the particles. Of these sixteen structures, one is spin-independent, six involve the spin of one of the particles, and the remaining nine involve both particle spins. Ten of the sixteen depend on the relative momenta of the particles. The addition of the spin degree of freedom opens up a large variety of possible new interactions to search for which might have escaped detection to date. Powerful astrophysical constraints on exotic spin-dependent couplings\,\cite{PhysRevD.51.1495, Raffelt1995b, PhysRevD.86.015001} exist from stellar energy-loss arguments, either alone or in combination with the very stringent laboratory limits on spin-independent couplings from gravitational experiments\,\cite{PhysRevD.88.031101}. However, a chameleon mechanism could in principle invalidate some of these astrophysical bounds while having a negligible effect in cooler, less dense lab environments\,\cite{Jain2006}, and the astrophysical bounds do not apply to axial-vector couplings\,\cite{Dob06}. These potential gaps in the astrophysical constraints, coupled with the intrinsic value of controlled laboratory experiments and the large range of theoretical ideas which can generate exotic spin-dependent couplings, has led to a growing number of searches for such effects in laboratory experiments\,\cite{Safronova2017}. 

Laboratory constraints on possible new interactions of mesoscopic range which depend on {\it both} the spin {\it and} the relative momentum are less common, because the polarized electrons or nucleons in most experiments employing macroscopic amounts of polarized matter typically possess $\langle\vec{p}\rangle=0$ in the lab frame. Some limits exist for spin-$0$ boson exchange\,\cite{PhysRevA.82.062714, PhysRevLett.112.091803} and spin-$1$ boson exchange\,\cite{PhysRevLett.110.082003, PhysRevD.88.031101, PhysRevLett.115.182001}. Spin and velocity-dependent interactions from spin-$1$ boson exchange can be generated by a light vector boson $X_{\mu}$ coupling to a fermion $\psi$ with a functional form of $\mathcal{L}_{I}=\bar{\psi}(g_{V}\gamma^{\mu}+g_{A}\gamma^{\mu}\gamma_{5})\psi X_{\mu}$, where $g_{V}$ and $g_{A}$ are the vector and axial couplings. In the nonrelativistic limit, this Lagrangian gives rise to two potentials of interest depending on both the spin and the relative momentum\,\cite{JPConfSer.340.012043}: one proportional to $g_{A}^{2}\vec{\sigma}\cdot(\vec{v}\times\hat{r})$ and another proportional to $g_{V}g_{A}\vec{\sigma}\cdot\vec{v}$. As noted above, many theories beyond the Standard Model can give rise to such potentials. For example, spontaneous symmetry breaking in the Standard Model with two or more Higgs doublets with one doublet responsible for generating the up quark masses and the other generating the down quark masses can possess an extra U(1) symmetry generator distinct from those which generate $B$, $L$, and weak hypercharge $Y$. The most general U(1) generator in this case is some linear combination $F=aB + bL +cY + dF_{ax}$ of $B$, $L$, $Y$, and an extra axial U(1) generator $F_{ax}$ acting on quark and lepton fields, with the values of the constants $a,b,c,d$ depending on the details of the theory. The new vector boson associated with this axial generator can give rise to $\mathcal{L}_{I}$ above\,\cite{FAYET1990743}. 

Piegsa and Pignol\,\cite{PhysRevLett.108.181801} recently reported improved constraints on the product of axial vector couplings $g_{A}^{2}$. They sought a potential of the form

\begin{equation}
V(\vec{r},\vec{v})\equiv V_5=\frac{g_A^2}{4\pi m}\frac{e^{-m_0r}}{r}\left(
\frac{1}{r}+\frac{1}{\lambda_c}
\right)\vec{\sigma}\cdot(\vec{v}\times\hat{r})
\label{eq:v5}
\end{equation}
and we will refer to this potential as $V_5$ in this paper. Here $m$ is the neutron mass, $m_{0}$ is the exchange boson mass, and $\lambda_c=1/m_{0}$ is the Yukawa range given by the Compton wavelength of the exchange boson. Polarized slow neutrons which pass near the surface of a plane of unpolarized bulk material in the presence of such a potential experience a phase shift which was sought in this experiment using Ramsey's well-known technique of separated oscillating fields\,\cite{PhysRev.78.695}. 

In this paper we report a more sensitive search for $V_{5}$ using polarized slow neutron spin rotation. Our idea in this experiment was to improve upon the Piegsa and Pignol work by simply increasing the total number of neutrons used to probe the possible spin dependent effect and to employ spin rotation as the measurement method rather than Ramsey spectroscopy. Vertically polarized neutrons are rotated about the transverse axis and thus \enquote{tipped} forward or backward under the influence of $V_5$ for horizontally stacked target masses. Slow neutron polarimetry has been used to search for parity violation in neutron spin rotation in $^{4}$He\,\cite{PhysRevC.83.022501} and to constrain possible exotic parity-odd couplings of the neutron\,\cite{PhysRevLett.110.082003} and polarized neutron couplings to in-matter gravitational torsion\,\cite{PhysLettB.730.353,PhysLettB.744.415} and in-matter nonmetricity\,\cite{PhysLettB.772.865}.

\section{Spin Rotation Experiment and Apparatus}

The experiment was conducted on the FP12 neutron beamline at the Los Alamos Neutron Scattering Center\,\cite{LANSCE} at Los Alamos National Lab. Bursts of 800 MeV protons from the proton linac and storage ring strike a tungsten spallation target at $20\,\mbox{Hz}$\,\cite{PSR}, thereby producing neutron bursts in 50 msec long frames. A liquid hydrogen moderator produces an approximately Maxwellian neutron energy spectrum with an effective temperature of about $40$\,K and a peak intensity at an energy of $3\,\mbox{meV}$\,\cite{seo2005CP769}. The neutrons pass down a $17~\mbox{m}$ long,  $9.5~\mbox{cm} \times 9.5\,\mbox{cm}$ cross sectional area supermirror neutron guide with $m=3$. The resulting beam divergence $21~\mbox{m}$ downstream at the entrance to the experimental cave was measured in Ref.\,\cite{seo2005CP769} and is in agreement with the expected phase space acceptance of the guide.  

\begin{figure*}
\centering
\includegraphics[scale=0.5]{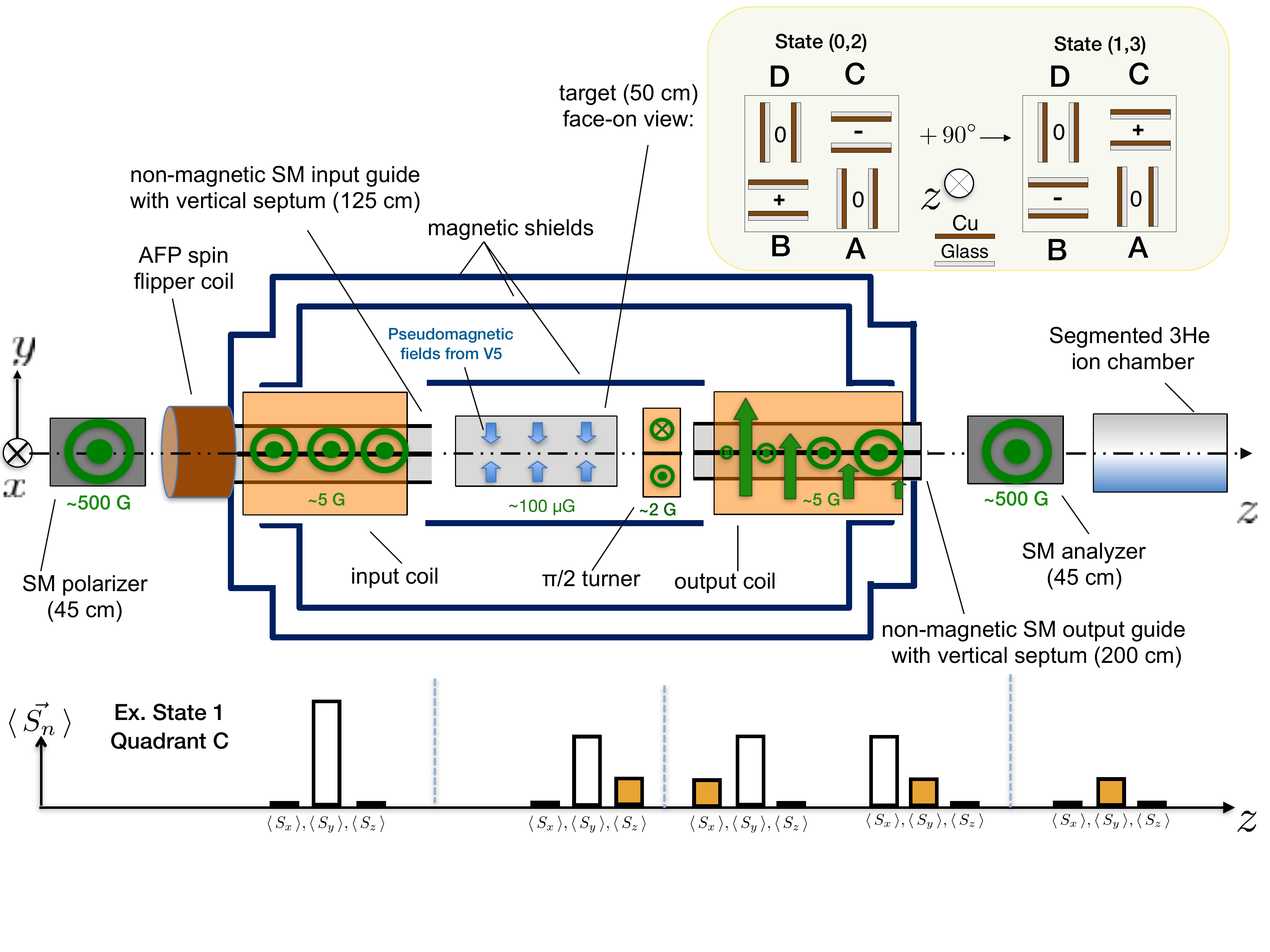}
\caption{A bird's eye view schematic of our neutron polarimeter. Neutrons are incident from the left. The vertical septum coincides with the z-axis. The magnetic field experienced by neutrons along the flight path is illustrated on the field components themselves, while the neutron's spin expectation value is depicted in the bottom plot at various stages along the polarimeter. A target state is assumed such that the neutron experiences a \enquote{forward tilt} due to the exotic potential, indicated by a solid bar, and the spin flipper is not energized.}
\label{fig:polarim}
\end{figure*}

Our apparatus is the neutron equivalent of a crossed polarizer/analyzer pair of the type familiar from light optics and is schematically pictured in Figure\,\ref{fig:polarim}. The design and operation of this neutron polarimeter have been discussed in great detail in the literature\,\cite{RSI.86.055101} so we confine ourselves to a very brief description here. Neutrons are polarized using a supermirror neutron polarizer and are adiabatically transported by an input coil lined with a nonmagnetic supermirror neutron guide. The end of the input coil is engineered to produce a nonadiabatic transition for the neutron spin as it is launched into a magnetically shielded region. In this low field region, the neutrons interact with a target via $V_5$ and accumulate a longitudinal polarization component. The neutron polarization is then rotated by $\pi/2$ radians by a so-called \enquote{$\pi/2$ coil} so that the polarization component of physical interest is captured by a nonadiabatic transition into a horizontal output coil field and internal nonmagnetic neutron guide.  This output coil and guide transports the neutrons to a polarization analyzer, and the transmitted neutrons are detected in an ionization chamber operated in current mode.  
The output coil is designed to produce a horizontal field at its entrance which rotates to the vertical direction by the end of the coil.  This field adiabatically rotates the horizontal component of the $V_5$ rotated spin to the vertical direction so it can be analyzed by the vertical orientation of the supermirror analyzer.  By flipping the horizontal component of the output coil entrance field, we alternately analyze the two components of the spin state.

We modified the polarimeter as described in\,\cite{RSI.86.055101}, which was designed to search for parity-odd rotations of the neutron spin about the neutron momentum, to search for neutron spin rotation about an axis normal to the neutron momentum. We used $m=2$ input and output guides constructed from nonmagnetic NiMo/Ti multilayers to transport most of the phase space of the neutron beam and preserve the neutron polarization. The neutron beam is split horizontally in the input guide into two regions by a central vertical septum coated on both sides with supermirror so that signals from the two halves of the beam can be used to cancel common mode systematic effects in the measurement and to make the measurement insensitive to possible nonstatistical noise from the neutron source. The input and output coils consist of wires woven into grooves etched into hollow hexagonal plastic extrusions with a rectangular cavity in the center to fit the input guide. The locations of the wires were determined by working backwards from the magnetic field shape required to realize the adiabatic and nonadiabatic neutron spin transitions on either end\,\cite{NIMA.854.127}. Finally the magnetic field in the precession coil of the polarimeter was operated with a lower current to realize a $\pi/2$ precession of the neutron spin about a vertical axis rather than a $\pi$ precession. This coil is composed of two rectangular solenoid coils joined together using two half-toroid coils on the top and bottom such that the field generated when energized is a continuous loop producing opposing vertical magnetic fields for the left and right halves of the beam. This field precesses the newly accumulated longitudinal spin component about the vertical ($y$) axis from $V_{5}$ to along the transverse  ($x$) axis, which is then in the correct position to be analyzed downstream by the output guide and coil and the polarization analyzer.  We applied transverse magnetic fields of $10\,\mbox{mG}$ in the target region to confirm that the polarimeter and  $\pi/2$ coil functioned as described.  

The target design for this experiment, described in detail in\,\cite{NIMA.2017}, is also qualitatively different from that used to search for parity-odd neutron spin rotation. In order to maximize the total number of neutron-atom interactions with the target while remaining sensitive in the mesoscopic length region of scientific interest, we designed a target using multiple flat plates containing a large mass density gradient across the gaps traversed by the polarized neutrons. The test masses were arranged in four quadrant regions each containing eight open channels for the neutrons separated by two plate thicknesses. The test masses on either side of the channels are composed of Cu ($N_{\text{Cu}} = 5.4\times 10^{24}/\mbox{cm}^3$), and float glass ($N_{\text{gl}} = 1.6\times 10^{24}/\mbox{cm}^3$). Test masses with a difference in mass density produce a nonzero $V_5$ between plates. The gaps in the quadrants are oriented so that two of the quadrants are sensitive to $V_{5}$ and the other two are insensitive. The ion chamber after the polarization analyzer possesses a matching set of quadrants, each with four charge collection planes along the neutron beam, for a total of sixteen independent charge collecting regions. The ion chamber was shown to produce a Poisson-like distribution when tested at the LENS neutron source at Indiana University by operating in pulse counting mode. Contributions to the uncertainty above $\sqrt{N}$ due to current mode operation increased the statistical fluctuation to $1.1\sqrt{N}$ \cite{Penn}. To reduce possible systematic errors from space-dependent nonuniformities in the background magnetic field as well as possible differences in target plate properties (flatness, thickness, etc.), it was crucial to have a mechanism to rotate the target in $90^\circ$ increments to allow neutrons to sample the same region of space with different plates in the opposite orientation so the $V_5$ rotation would change sign but magnetic rotations would not. Furthermore, by reversing the direction of the mass gradient from quadrant to quadrant we also reverse the sign of $V_5$ to allow comparison of rotations from different quadrants at taken at the same time. A Geneva drive mechanism translated continuous rotation from a rotating cam into an intermittent rotary motion like in a mechanical clock and was driven by an air motor located outside of the magnetic shielding to minimize stray magnetic fields in the target region. The flow was controlled by the Data Acquisition System (DAQ) via an analog relay actuated valve. Target state rotations took $2\,$seconds to complete, and an optical flag confirmed that the target reached each desired state.  

\section{Data Acquisition and Apparatus Characterization}
	
The current integrator output voltages from each of the sixteen ion chamber plates given by $V(t)=-I_0t/C$ are reset to $0$ volts near the end of the neutron frame by sending a pulse to the integrator's reset input. The DAQ  controlled the sequencing of polarimeter states by energizing the various field producing coils, operating the air motor used for target rotation, reading the optical signals to confirm the target state, measuring the neutron beam intensity by reading the voltage output from each of the sixteen current integrators, and monitoring the internal magnetic fields and currents supplied to the coils. We used a National Instruments PCIe-6363 X Series Multifunction DAQ card installed on a PC running Windows 7 with 32 analog input channels, 48 digital input/output channels, and 4 analog output channels. The ADC bit resolution made a negligible contribution to the signal noise. The DAQ was triggered by a pulse received at our flight path whenever the LANSCE proton beam struck the spallation source to generate neutrons, hereafter referred to as the \enquote{$T_0$} pulse.

Residual magnetization in the target plate masses could potentially produce a neutron spin rotation that would mimic or hide the effect of the spin dependent interaction of interest. Therefore the magnetization was scanned by small fluxgate magnetometers both before and after the experiment. We saw no evidence for the presence of any residual fields at the 10$\mu$G level at a distance of 3\,mm from the plates. The resulting upper bound on systematic errors in our measurement from this effect is below $2\times 10^{-8}$\,rad.

The efficiencies and the spatial uniformity of the supermirror neutron polarizer and analyzer were measured at the Low Energy Neutron Source (LENS) at Indiana University\,\cite{LEUSCHNER2007956, LAVELLE2008324}. We found each device produced $\mathcal{P}>0.95$ and $\mathcal{A}>0.95$ over the cold neutron spectrum, consistent with simulations and the data provided by the manufacturer. The value of the total polarizer-analyzer product, $\mathcal{PA}$, of our polarimeter as assembled on the FP12 beam was measured by applying a uniform horizontal transverse magnetic field in the target region and measuring the count rate asymmetry as a function of current in the coil. By locating the field which maximizes the asymmetry, we isolate the $\mathcal{PA}$ product. The proper orientation of the ion chamber quadrants with the input/output guides was realized by rotating the cylindrical ion chamber about the neutron beam axis and equalizing the signals from the quadrants. The optimal current for the $\pi/2$ coil was chosen using a longitudinal solenoid in the target region which produced a known field integral and therefore a known transverse neutron spin rotation for a given neutron speed. By adjusting the current to minimize the measured transverse rotation angle after passing through the $\pi/2$ coil we found an optimal value for the current. Neutrons which are over- or under-rotated at this value reduce the efficiency of the polarimeter which is reflected in the measured value of the so called \enquote{$\mathcal{PA}$ product} described in the next section.

The magnetic fields of all coils used in the experiment were mapped over the volume occupied by the neutron beam to ensure proper neutron spin transport. The magnetic field was not actively stabilized during data aquisition. Fields inside the target region were a few hundred $\mu$G. Environmental magnetic field drifts at the level of a few $\mu$G over the timescales for changes in the target and apparatus states. The magnetic shields were typically degaussed a few times/day. We employed four functional single-axis fluxgate magnetometers located several cm above the target and neutron beam (one fluxgate stopped working properly during the run). Three were located about 10\,cm directly above the target with orientations alternating along the beam and transverse to the beam, and a fourth fluxgate was located downstream from the target. The fields during a typical run varied by a couple of $\mu$G, with slightly larger amplitudes closer to the exit of the magnetic shield as one would expect. The fields in all fluxgates averaged over the runs used to extract the neutron spin rotation angle were consistent with zero within the 1 $\mu$G sensitivity of the fluxgates.

\section{Data Analysis and Results}

The data were taken over a series of repeated 32-minute sequences, which was the amount of time necessary for the apparatus to change to each unique field and target configuration state, while accumulating statistically interpretable data in each state. There were six $\pi/2$ coil states $(+,-,0,+,-,0)$ and eight target states $(0,1,2,3,4,5,6,7)$ where $(4,5,6,7)$ are identical to $(0,1,2,3)$.
The target was designed such that in any of the four possible target states two diagonally opposed quadrants would not be sensitive to $V_5$ while the remaining two would produce $V_5$ rotations of opposite signs from opposite mass-gradients.  Since the $\pi/2$ coil generates fields of opposing signs in both halves of the beam, the rotations from $V_5$-sensitive quadrants will have the same sign after the analyzer. Rotating the target into different states reverses the mass gradient and thus the signs of the $V_5$ rotations in each quadrant without affecting the signs of rotations from magnetic fields. In the following treatment we will assume uniform background fields. The general case has been treated in Ref.\,\cite{NIMA.2017} where it is shown that the effects of field nonuniformities are shown to contribute a systematic error only when target-state subtraction is ineffective due to target misalignments or differences in the reflectivities of the copper or glass target plates. Estimates of these systematics are listed in Table\,\ref{tbl:sys}.

The rotations in each quadrant (see image of target in  Fig.\,\ref{fig:polarim}\,: A lower right, B lower left, C upper right, D upper left) in terms of background transverse magnetic fields, $B_0^T$, and possible rotations from $V_5$ are given by

\begin{align}
	\phi_A & = \phi_{B_0^T}\label{eq:phi1}\\
	\phi_B & = \pm\phi_{V_5}-\phi_{B_0^T}\\
	\phi_C & = \pm\phi_{V_5}+\phi_{B_0^T}\\
	\phi_D & = -\phi_{B_0^T}
\label{eq:phis}
\end{align}

The positive and negative signs in Eq.\,(\ref{eq:phi1})\,-\,Eq.\,(\ref{eq:phis}) describe the $V_5$ rotation in target states (0,\,2) and (1,\,3), respectively.  Therefore, we may isolate the contribution to the total spin rotation from $V_5$ in two ways:

\begin{enumerate}
\item Average asymmetries formed in quadrants B and C ($V_5$ non-zero) in any target state and average over all target states, i.e.
\begin{align}
2\phi_{V_5}&=\frac{(\phi_ B + \phi _C)_0}{4}-\frac{(\phi _B + \phi _C)_1}{4}\label{eq:quadav}\\ &+\frac{(\phi _B + \phi _C)_2}{4}-\frac{(\phi _B + \phi _C)_3}{4}
\nonumber
\end{align}
or 
\item Subtract asymmetries formed in the same quadrant but in different target states and then take the average over quadrants, i.e.
\begin{align} 2\phi_{V_5}&=
\frac{(\phi _{B0} - \phi _{B1})}{4}+\frac{(\phi _{B2} - \phi _{B3})}{4}\label{eq:stateav}\\ 
&+\frac{(\phi _{C0} - \phi _{C1})}{4}+\frac{(\phi _{C2} - \phi _{C3})}{4}\nonumber
\end{align}
\end{enumerate}

We found that the results from these two methods were consistent as expected since we use the same steps between the two methods just reversed in order. In the data analysis, we use Eq.\,(\ref{eq:stateav}) to isolate the $V_5$ rotation. Note that we only need to consider the $B$ and $C$ quadrants when isolating the $V_5$ rotation; however, any background fields that remain after the subtraction in Eq.\,(\ref{eq:stateav}) must be accounted for in the final result. Such rotations could arise from fluctuations in beam intensity and target state-dependent effects. The residual rotation $\phi_{res}$ is given by 

\begin{align}
2\phi_{res}&=\frac{(\phi _{A0} - \phi _{A1})}{4}+\frac{(\phi _{A2} - \phi _{A3})}{4}\label{eq:residrot}\\ &+\frac{(\phi _{D0} - \phi _{D1})}{4}+\frac{(\phi _{D2} - \phi _{D3})}{4}.\nonumber
\end{align}

Therefore, the quantity of interest which fully isolates the effect of $V_5$, $\phi^{\prime}$, is computed as

\begin{equation}
\phi^{\prime} = \phi_{V_5} - \phi_{res}.
\label{eq:phip}
\end{equation}

  Neutron spin rotations in this measurement are extracted from neutron intensity asymmetries. These asymmetries, shown in Eq.\,(\ref{eq:asymsA}), are formed by comparing intensities of the different output coil states OC$\pm$ every $2\,$seconds as it was important to change this state as often as possible to reduce effects primarily from possible time dependent ambient magnetic fields. The current in the $\pi/2$ coil cycled from $\pi/2(+):I=90\,\mbox{mA}$ to $\pi/2(0):I=0\,\mbox{mA}$ to $\pi/2(-):I=-90\,\mbox{mA}$. For each $\pi/2$ coil state the output coil flipped twenty times. Changing the $\pi/2(\pm)$ current simply reverse the sign of all measured spin rotation angles so that averaging over these states can reduce systematic effects from possible nonuniformities in the $\pi/2$ coil magnetic field. 

To form the asymmetry from the neutron intensity measurements, we first convert the charge integrated over one $T_0$ pulse to the number of incident neutrons using the known number of ion pairs per neutron and the charge to voltage ratio for each collection plate deduced by sending in known currents to the plates directly \cite{CCH}. An asymmetry is computed at the end of 40 $T_0$ pulses, or 2 seconds. After ten asymmetries are formed in each $\pi/2$ coil state an average asymmetry is computed where the error is given as the deviation from the mean of the number of incident neutrons. The neutron counts were normalized by the proton beam current in each pulse to eliminate any effects from slow drifts in the beam intensity. The normalization adds a negligible contribution to our statistical uncertainty. The expression for the total spin rotation angle $\phi$ measured in each quadrant is given in terms of the measured asymmetry, $A_L$ , as

\begin{align}
	A_L = \mathcal{PA} \text{sin}(\phi) &= \frac{N^+-N^-}{N^++N^-}
	\label{eq:asymsA}\\
	\leftrightarrow \phi &\simeq \frac{1}{\mathcal{PA}}\frac{N^+-N^-}{N^++N^-}
\end{align}

\begin{figure}
\centering
\includegraphics[scale=0.47]{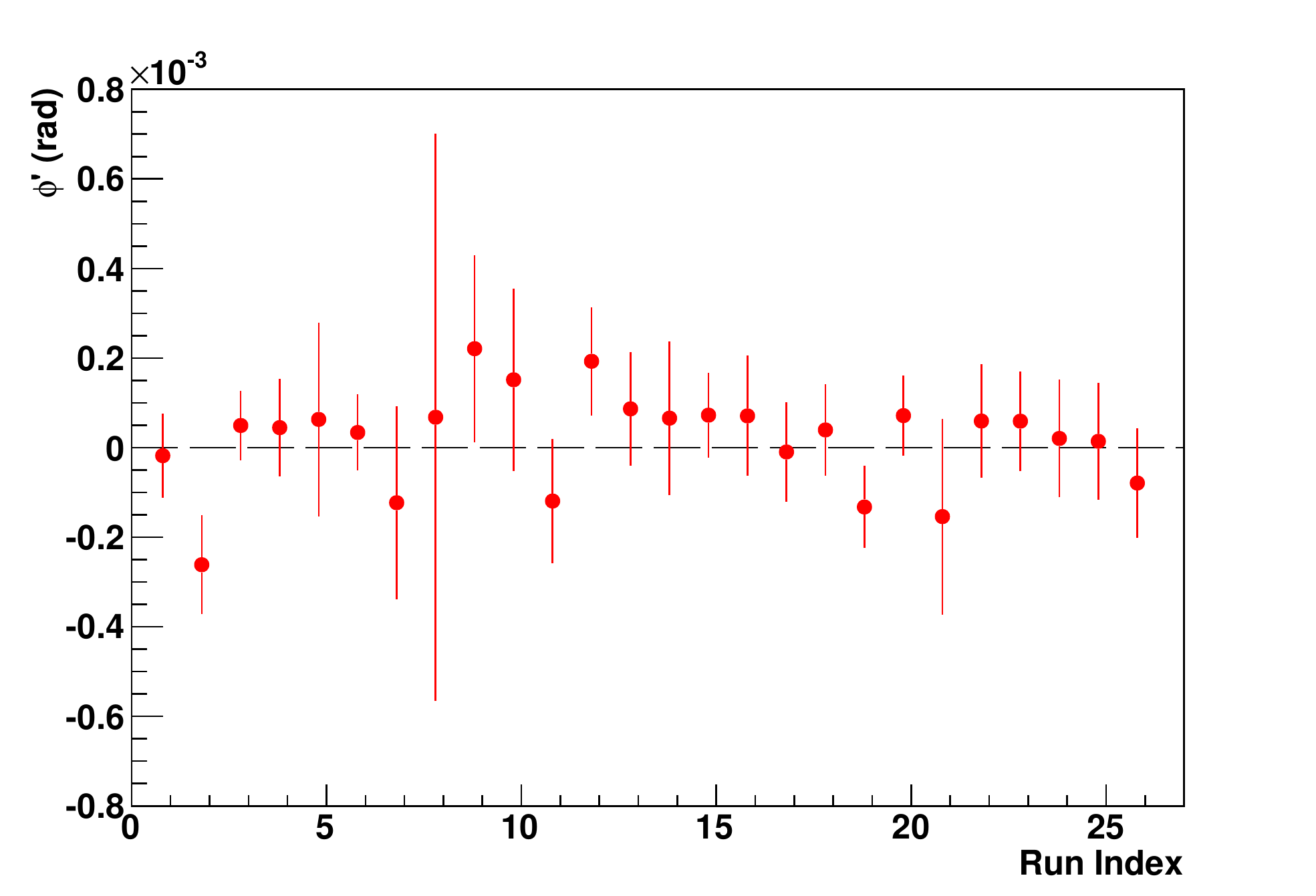}
\caption{26 production runs averaged over $\pi/2$ coil states and subtracted to isolate quadrant-dependent physics. The weighted mean for this set is $\bar{\phi'}=(1.4\pm 2.3)\times 10^{-5}~\mbox{rad}$ with $\chi^2$/NDF = 0.7. The vertical errors include any deviations from the mean of the number of incident neutrons. The dashed line is at zero across the whole range.}
\label{fig:dataprod}
\end{figure}

Once an asymmetry is formed in the analysis it must be divided by the corresponding quadrant-dependent $\mathcal{PA}$ product, whose error makes a negligible contribution to our overall measured uncertainty. For each experimental run a value of $\phi_{V}^\prime$ and its error is extracted for each $\pi/2$ coil state in the four quadrants, a total of twelve values for each run. A total of 26 runs were combined to produce Figure\,\ref{fig:dataprod}. Using the data points and their errors we find the weighted mean $\bar{\phi^\prime}$ equal to $(2.8\pm 4.6)\times 10^{-5}$\,rad/m. The target state and quadrant subtraction reduction scheme reduces the size of the raw spin rotation angles from residual internal magnetic fields by a factor of $\sim 10^3$.

\section{Systematic Effects}

Table\,\ref{tbl:sys} shows our estimates for the sizes of the various forms of systematic error for our target design and experimental upper bounds. The great majority of the systematic effects, whose various sources are described in detail in Ref.\,\cite{NIMA.2017}, are associated with residual magnetic fields coupled with various types of apparatus or beam nonuniformities. The dominant term in Table\,\ref{tbl:sys} comes from measurements using data taken with the $\pi/2(0)$ data combined with our knowledge of the internal magnetic field in the apparatus, which we measured continuously using fluxgate magnetometers.

The $\pi/2(0)$ states do not project rotations about the transverse axis onto the $x-y$ plane and therefore asymmetries formed in this state are not sensitive to systematic effects from rotations about transverse fields in the target region.  However, they do serve as a check on the size of systematic effects from rotations about longitudinal fields throughout the low-field region.  When in the $\pi/2(1)$ ($\pi$-coil on) states only the portion of longitudinal fields in the short distance between the $\pi$-coil and output coil cause rotations that are analyzed by the polarimeter.  Since the ratio of distance after the $\pi$-coil to the total length of the low-field region is about a factor of $0.25$, we can scale the  $[-4.0\pm 2.8(stat)]\times 10^{-5}$\,rad/m result from the $\pi/2(0)$ data down by a factor of four for the $\pi/2(1)$  case.  This places an experimental upper limit on systematic errors from rotations about longitudinal fields while in the $\pi/2(1)$ states of $1\times 10^{-5}$\,rad/m.

The largest sources of systematic errors described in Ref. [33] are due to target magnetic impurities, target misalignments and differences in target-plate reflectivities, which cause an incomplete subtraction of magnetic field rotations after the cancellation procedures described above.  The first does not scale with ambient field, while the others do and thus are expected to be larger given the difference between the fields assumed in the estimates of Ref.\,\cite{NIMA.2017} an the fields measured in our apparatus.  As already described, the $\pi/2(0)$ data place a limit on the longitudinal portion of the sum of the latter two systematic effects.  Fluxgate and field map measurements suggest that transverse fields are of roughly the same size  or smaller than the longitudinal fields.  Given the 3 times longer distance before the $\pi/2$-coil, we estimate that the systematics from the transverse-field portion is 3 times larger than the longitudinal portion. This leads to an experimental upper limit for the systematic errors due to target misalignment and reflectivity differences of $4\times 10^{-5}$\,rad/m. This is the dominant source of systematic error in the measurement.

Table\,\ref{tbl:sys} lists our estimates for the other sources of systematic error not constrained by the $\pi/2(0)$ data. To scale the systematics described in Ref.\,\cite{NIMA.2017} for the case of the actual experiment, we need to know the size of the magnetic field within the target region. For this purpose we used magnetic field measurements from fluxgates placed within the magnetic shielding and just above the target. We also used our simulations combined with field maps of the apparatus to determine expected rotation values from magnetic field distributions and compare them with the observed raw spin rotation angles. The simulations propagate the spins through the field maps of residual fields from the input and output coils, the target region, and the $\pi/2$-coil. Both methods indicate the presence of internal magnetic fields of about 2\,mG in the target region. This relatively large internal magnetic field was due to space constraints on the FP12 beamline which precluded the use of the full complement of magnetic shielding of the apparatus and can therefore be greatly improved in future work. The measured size of this upper bound is slightly better than the somewhat pessimistic estimates presented in Ref.\,\cite{NIMA.2017} if the latter are scaled up to reflect the magnetic fields in the target region measured during the experiment.

Given the various subtractions discussed here and in Ref.\,\cite{NIMA.2017} which eliminate systematic effects from neutron cross talk between quadrants, the dominant systematic error arises from target misalignment and reflectivity differences in the target plates, which has an upper limit of $4 \times 10^{-5}$ rad/m. This result is derived from a combination of the $\pi/2(0)$ data, fluxgate measurements, and Monte Carlo simulations.

\indent It is important to make sure that there is no false systematic effect from crosstalk of signals in the data acquisition system. The size of these voltage fluctuations were investigated directly during the LANSCE run by closing the neutron shutter and running the DAQ as if it was taking production data. This includes rotating the target, flipping field producing coils, etc. These runs were interleaved with the production data runs and saw no effect at the $10^{-8}$ level. Further, type 1 ceramic capacitors were used to store the integrated charge which have a very low temperature coefficient $< 1\times 10^{-4}$/K. Therefore in our experimental area which was temperature controlled to within 2K/day, asymmetries acquired on the time scale of 1\,s resulting from the temperature dependence of the capacitance amount to $< 2\times 10^{-9}\times V_{\text{max}}$ where $V_{\text{max}}$ is the max voltage of the preamp over an output coil flip,  which is negligible compared to our asymmetry signal.

\begin{table}[h]
\caption{A list of systematic effects in our search for $V_{5}$ using a slow neutron polarimeter. These estimates all hold for the internal  magnetic fields of 2 mG measured in the apparatus during the experiment using fluxgate magnetometers. We have included all systematic errors associated with analysis after both modes of target cancellation (diagonal averaging followed by $90^\circ$ target rotation). Systematics associated with differences in target plate reflectivities and misalignment are defined as \enquote{target variability}. All of the dominant sources of systematic error on this list scale with the size of these residual internal fields. Systematic errors due to target misalignment and reflectivity differences are constrained by $\pi/2(0)$ data.}
\begin{center}
\label{tbl:sys}
\begin{tabularx}{0.48\textwidth}{X|l}
\hline
Source of systematic & Uncertainty (rad/m) \\
\hline
Small angle scattering & $< 4 \times 10^{-6}$ \\
Diamagnetism & $< 4 \times 10^{-8}$ \\
Neutron-atom\,spin-orbit scattering & $1 \times10^{-8}$ \\
Target magnetic impurities & $<4\times10^{-6}$ \\
Target variability & $<4\times 10^{-5}$ \\
Electronic crosstalk & $<2\times 10^{-8}$ \\
\hline
Total &  $<4\times10^{-5}$ \\
\hline
\end{tabularx}
\end{center}
\end{table}

\section{Constraints on $g_{A}^{2}$}
\begin{figure}
\centering
\includegraphics[scale=0.45]{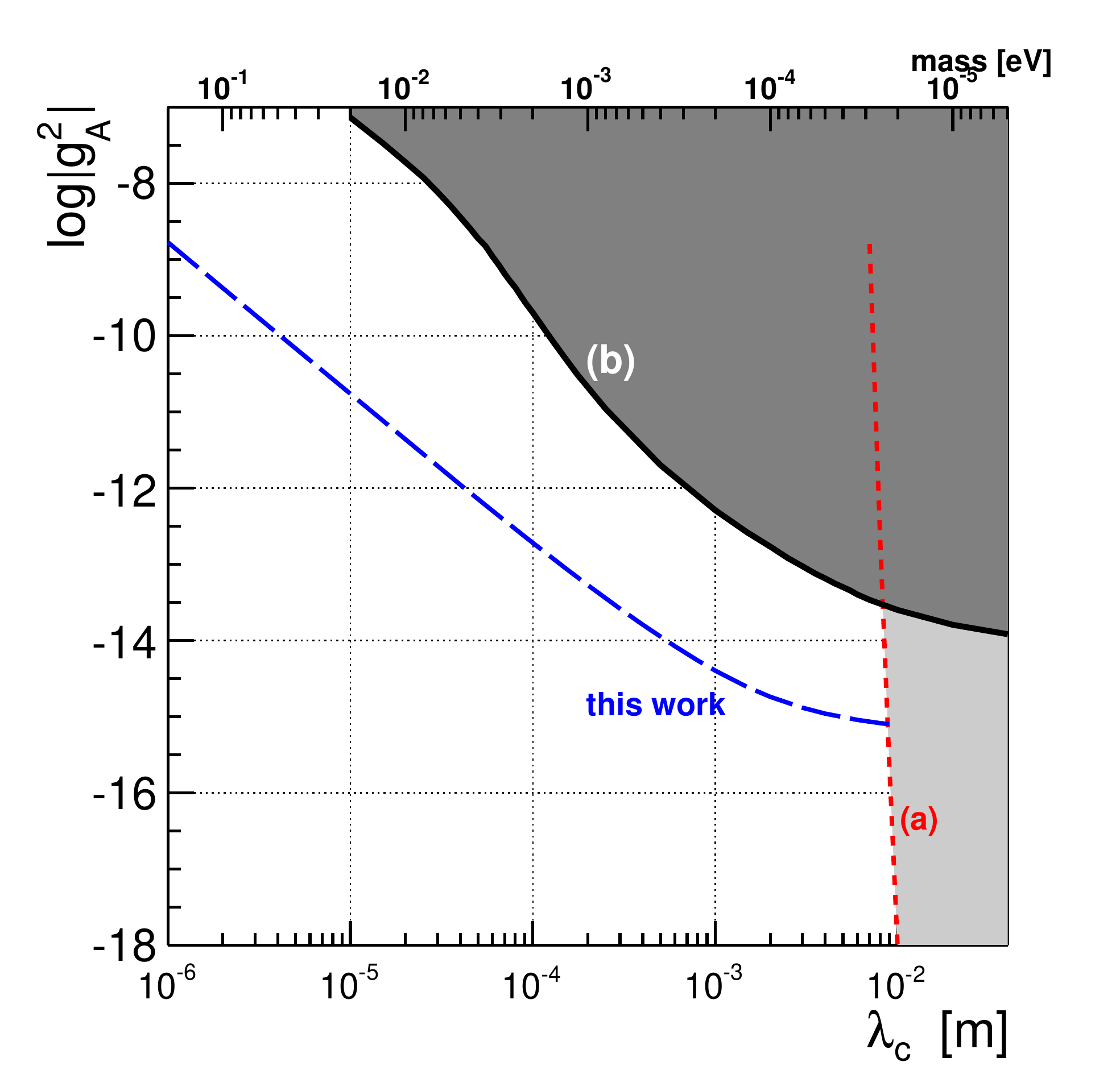}
\caption{$g_A^2$ as a function of $\lambda_c$ from our experiment (dashed-blue) compared with constraints from a neutron measurement using Ramsey spectroscopy (a)\,\cite{PhysRevLett.108.181801} and from K-$^3$He comagnetometry (b)\,\cite{PhysRevLett.103.261801}. The final $g_A^2$ limit includes both statistical and systematic uncertainties.}
\label{fig:ppimposed}
\end{figure}

To determine the sensitivity of the apparatus to $V_{5}$ we conducted a Monte Carlo simulation which integrated the expression in Eq.\,\ref{eq:v5} over the geometry of the target using the neutron energy spectrum on FP12 at LANSCE to give the relationship between the parameters in $V_{5}$ and the spin rotation angle $\phi$.  As discussed in Ref.\,\cite{NIMA.2017} the Monte Carlo simulations of the sensitivity to $g_A^2$ agree with an analytic expression assuming infinitely thick slabs for small Yukawa length scales but are nearly a factor of 10 less sensitive at $\lambda_c =3$\,mm due to the finite thickness of the slabs. From the measured asymmetry $\phi^{\prime} = [2.8 \pm 4.6 (stat.) \pm 4.0(sys.)] \times 10^{-5}$\,rad/m, which is consistent with zero, we derived an exclusion plot in the $g_{A}^{2}$, $\lambda_c$ parameter space shown in Figure\,\ref{fig:ppimposed}. This constraint improves on the previous upper bounds on $g_{A}^{2}$  by 2\,-\,4 orders of magnitude for $\lambda_c$ between 1\,cm and 1\,$\mu$m. 

\section{Conclusion}

We looked for a spin-dependent interaction between polarized neutrons and matter from an exotic light vector boson $X_{\mu}$ coupling to a fermion $\psi$ with form $\mathcal{L}_{I}=\bar{\psi}(g_{V}\gamma^{\mu}+g_{A}\gamma^{\mu}\gamma_{5})\psi X_{\mu}$, where $g_{V}$ and $g_{A}$ are the vector and axial couplings and $\lambda_c=1/m_{B}$ is the interaction range from a boson exchange of mass $m_{B}$. Our result of $\phi'_{V_5} = [2.8 \pm 4.6 (stat.) \pm 4.0(sys.)] \times 10^{-5}$\,rad/m is consistent with zero. We interpret this result as setting an upper bound to the strength and range of $V_{5}$. For Yukawa ranges between 1\,cm and 1\,$\mu$m our limits are more stringent than the previous measurement by about a factor of 1000.  

Prospects for future improvement in the sensitivity of a $V_{5}$ search are excellent. The data analyzed in this paper correspond to about one week of real-time running at LANSCE FP12. The intensity of cold neutron beams such as the NG-C beam at NIST and the PF1b beam at ILL is higher by about two orders of magnitude. Combined with a longer running time and the use of a denser target material such as tungsten in place of the copper used in this work, one can envision a further improvement in the statistical sensitivity to a $V_{5}$ potential in future measurements of more than two orders of magnitude assuming that there are no limitations from unanticipated systematic errors. The internal magnetic fields ($\sim$mG) which determined the size of the systematic error can be reduced by at least three orders of magnitude to be negligible compared to the statistical error. Such an experiment would probe neutron axial couplings to matter through an exotic spin 1 boson exchange which are about $13$ orders of magnitude weaker than electromagnetism.

\section{Acknowledgments}

We gratefully acknowledge the local support of the LANSCE neutron facility at Los Alamos National Lab where this measurement was performed. Melvin Borrego and Kelly Knickerbocker provided outstanding support by going well out of their way to ensure that our experiment ran seamlessly. We also acknowledge the support of the LENS neutron source at Indiana University where the neutron polarimeters and ion chamber were tested and characterized. We would like to thank Dr. Florian Piegsa and Dr. Guillaume Pignol for providing us with their experimental data, allowing us to produce the exclusion plot in Fig.\,3. C. Haddock, E. Anderson, W. Fox, I. Francis, J. Fry, K. Korsak, W. M. Snow, K. Steffen, and J. Vanderwerp acknowledge support from US National Science Foundation grants PHY-1306942 and PHY-1614545 and from the Indiana University Center for Spacetime Symmetries. C. Haddock acknowledges support from the US Department of Energy SCGSR Fellowship and the Japan Society for the Promotion of Science Fellowship. M. Sarsour acknowledges support from US Department of Energy grant DE-SC0010443. L.~Barr\'{o}n-Palos and M.~Maldonado-Vel\'{a}zquez acknowledge suppport from PAPIIT-UNAM grants IN111913 and IG101016. W. M. Snow also acknowledges support from the Mainz Institute of Theoretical Physics (MITP) for its hospitality and its partial support during the completion of this work. H. Gardiner, D. Esposito, and J. Lieffers acknowledge support
from the NSF Research Experiences for Undergraduates program
NSF PHY-1156540.


\bibliography{F5PLBdraft_May29}

\end{document}